# Hardware implementation of auto-mutual information function for condition monitoring

Harun SILJAK[1,*], Abdulhamit SUBASI[2], Belle R. UPADHYAYA[3]

[1] Electrical and Electronics Engineering Department, International Burch University, Sarajevo, Bosnia and Herzegovina

[2] College of Engineering, Effat University, Jeddah, 21478, Saudi Arabia

[3] Nuclear Engineering Department, The University of Tennessee

Knoxville, TN 37996-2300, USA

*Correspondence: harun.siljak@ibu.edu.ba

**Abstract:** This study is aimed at showing the applicability of mutual information, namely auto-mutual information function for condition monitoring in electrical motors, through age detection in accelerated motor aging. Vibration data collected in artificial induction motor experiment is used for verification of both the original auto-mutual information function algorithm and its hardware implementation in Verilog, produced from an initial version made with MATLAB HDL (Hardware Description Language) Coder. A conceptual model for industry and education based on a field programmable logic array development board is developed and demonstrated on the auto-mutual information function example while suggesting other applications as well. It has also been shown that attractor reconstruction for the vibration data cannot be straightforward.

# 1. Introduction

The question of motor age identification using vibration signals has been studied thoroughly in the past [1,2]. Methods based on dynamical systems theory have been proposed [3], but they ask for more thorough insight, which is provided in this paper. This study provides an analysis of dynamical system properties of artificial motor aging vibration data, a novel method for age detection and an implementation of this method in an innovative conceptual model.

Regarding realization, the idea of condition monitoring algorithm implementation in hardware is not new [4-6]. There have been efforts to implement some new and specific algorithms in hardware, but some classical algorithms have been neglected as their applicability to condition monitoring was not acknowledged. Implementations so far, as in general case of condition monitoring as well, have been focused on machine learning [4], while the work presented here will focus on a signal processing technique from the dynamical systems theory.

In work presented here, we introduce a novel framework for condition monitoring prototyping and training, which we use as the platform for a hardware implementation of auto-mutual information function calculator. While it is usually used in dynamical systems theory for attractor reconstruction, here we show that the auto-mutual information function is a good indicator of motor state, when applied to its vibration signals.

In the second section of the paper, we present the fundamental concepts our implementation relies on: the auto-mutual information function, hardware description and the problem of the artificial motor aging. While describing the listed concepts, we also discuss the way we implement them. In the third section, we show the results of our proposed solution applied to two fundamentally different problems of condition monitoring, artificial and non-artificial motor aging. Finally, we discuss the results obtained before drawing conclusions and implications for future work.

## 2. Materials and methods

### 2.1. Auto-mutual Information Function

To calculate Lyapunov exponents which are suggested as a feature useful for motor condition determination in earlier work [3], the attractor has to be reconstructed. Since in general, we do not have more than one or at most two time series from the process (and the underlying dynamics have higher dimensions), other coordinates in phase space need to be generated. A common approach is one using a delayed version of the existing coordinates and applying the Takens theorem [7]. This method depends heavily on two parameters, the delay time and the embedding dimension.

There is no optimal algorithm for these parameters, but some plausible approaches were introduced. For example, the delay time can be determined using auto-mutual information function [8], while a false neighbor-like approach can be used for embedding [9]. After this attractor reconstruction, the Lyapunov exponent can be calculated. The most used algorithm for that is one proposed in [10], but others have been proposed as well, such as Sato's [11], which is the algorithm for Lyapunov exponent calculation in MATLAB compatible OpenTSTOOL software [12] based on non-linear time series algorithms and methods presented in [13].

As the auto-mutual information function (AMIF) is going to be used extensively in this letter, it will be defined here.

For systems $S$ and $Q$ with discrete states $s_1, s_2, \ldots, s_n$ and $q_1, q_2, \ldots, q_m$ with respective probabilities $P_s(s_1), \ldots, P_s(s_n)$ and $P_q(q_1), \ldots, P_q(q_m)$ the mutual information function

$I(Q,S)$ is the number of bits of $q$ that can be predicted on average given a measurement of s:

$$I(Q,S) = \sum\sum P_{sq}(s_i q_j) \log \frac{P_{sq}(s_i q_j)}{P_s(s_i) P_q(q_j)} \quad (1)$$

where $P_{sq}(s_i,q_j)$ is the probability that $s=s_i$ and $q=q_j$. In order to apply this in experimental data, $P_{sq}$ is to be estimated by partitioning of $S$-$Q$ plane into elements. $P_{sq}$ is represented by the ratio of the number of points in an element and the total number of points.

After obtaining the first minimum of the auto-mutual information function (seen in a graphical representation in Figure 1 for a motor vibration signal), its index can be used as the delay time for generation of other coordinates in the phase space. Cao's dimension estimation algorithm is applied to determine the number of such coordinates. Abscissa value where its characteristic $d_1$ drops for the first time denotes the embedding dimension. Furthermore, its $d_2$ characteristic is used for determinism checking: it is constant in case of pure stochastic signals. After these two parameters (delay time and embedding dimension) are acquired, the reconstruction is straightforward.

At that point, the Lyapunov exponents can be calculated by applying any of the proposed algorithms. Namely, the Sato's algorithm finds the Lyapunov exponent for the time series as the slope of the particular probability curve in its rising part.

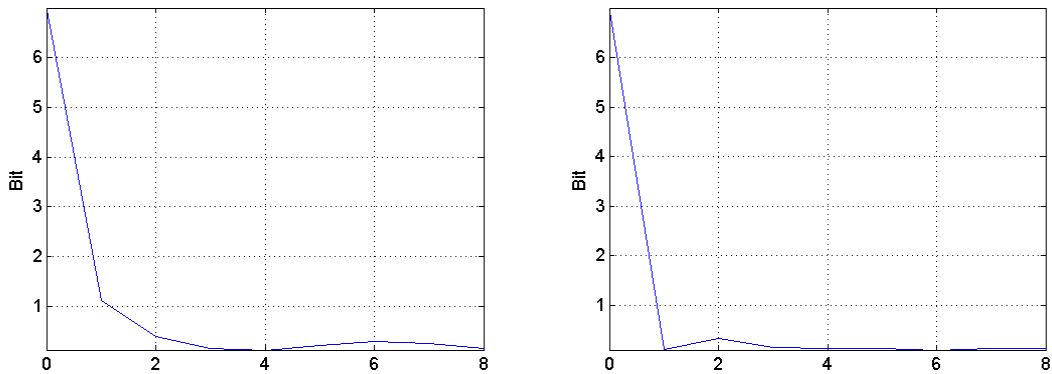

**Figure 1.** Auto-mutual information function for vibration series 0 (a) and 7 (b).

## 2.2. Hardware Description

Since this research is not conducted on a dedicated motor testbed, this practical implementation should have the possibility to be verified using the existing data records in a Hardware-in-loop (HIL) setup. Validation of condition monitoring methodologies in HIL setup has been done before, so it can be considered a reliable procedure [14].

While a generalized form of this device will be discussed in this subsection, a particular algorithm (auto-mutual information function) is chosen as a proof of concept to be implemented for demonstration purposes.

The system that is going to be developed within this study will have a physical system on chip (SoC) doing the condition monitoring and a virtual motor. Virtual motor's vibration is physical monitoring system's input, while the output of the physical system does not necessarily have to return to the virtual motor, which would make this more of a driven prototype than an HIL. As we are going to see, this system is, in fact, closer to the original concept of HIL.

Altera's FPGA (Field Programmable Gate Array) development board DE2i-150 is chosen for this project. This board's contents feature a combination of Altera Cyclone IV FPGA and an Intel Atom processor. This compact board, 26.7 cm × 17.6 cm × 4.2 cm weighing under 1 kg is easily transported, programmed through USB interface with a PC from Altera's Quartus II IDE. Altera's software enables easy SoC building with their own implementable processor Nios-II, and the Atom processor can run any operating system (Yocto Linux is pre-installed).[*]

The project outline is shown in Figure 2a, representing the following idea:

---

[*] Altera Inc. DE2i-150 Development Kit FPGA System User Manual. Terasic 2013.

1. The emulator subsystem can reproduce already recorded vibration data and send it to the FPGA design.
2. The FPGA design is an implementation of a vibration processing algorithm that accepts inputs either from the emulator subsystem or the real sensors.
3. Control input is a simple switching scheme in which the emulator subsystem and the FPGA algorithm implementation are informed of the working mode: emulator is commanded which of the predefined files to reproduce, and the FPGA part is informed whether to expect inputs from sensors or the emulator subsystem.
4. Sensor input is an accelerometer with analog-to-digital conversion suitable for real-life applications with induction motors.
5. Output can be any display representing the state of the motor as determined by the FPGA implemented algorithm.
6. The server is an external computer capable of commanding the emulator on its own.

To make this general design realistic, DE2i-150 board is used, and the design from Figure 2a becomes the concrete system in Figure 2b.

1. The emulator subsystem is C-code running on Intel Atom with Yocto OS communicating with FPGA through PCIe interface.
2. The FPGA design is placed on the Altera Cyclone IV.
3. Control input can be either the set of switches from the DE2i-150 board or the IR remote controller with DE2i-150's IR receiver.
4. Sensor input is DE2i-150's on-board three-axis accelerometer.

5. Output can be the DE2i-150's LCD or Ethernet (or Intel Wi-Fi) since it is relatively simple to make a web server on the FPGA part using Nios-II (an out-of-the-box web server solution is delivered with all DE2 boards).
6. The server is an external computer capable of commanding the emulator through SSH (Secure Shell). This can also be done using wireless, so the boards act as parts of a wireless sensor network.

It is worth noting that the emulator and the algorithm can coexist in a single chip: we use separate entities (FPGA and Intel Atom processor) for enhanced performance of the emulator, but it could have been on the FPGA together with the algorithm implementation, as a separate module connected to the same bus. Another convenient reason to use separate chips is keeping the emulator fixed (processor + dedicated memory) and non-volatile while changing the algorithm implementation all the time (FPGA).

A system like this one is another contribution in introducing HIL to curricula which is a growing trend for over a decade now [15,16] and enabling dynamical learning in areas of embedded systems design, condition monitoring, signal processing and even web programming if everything is done through (web) servers.

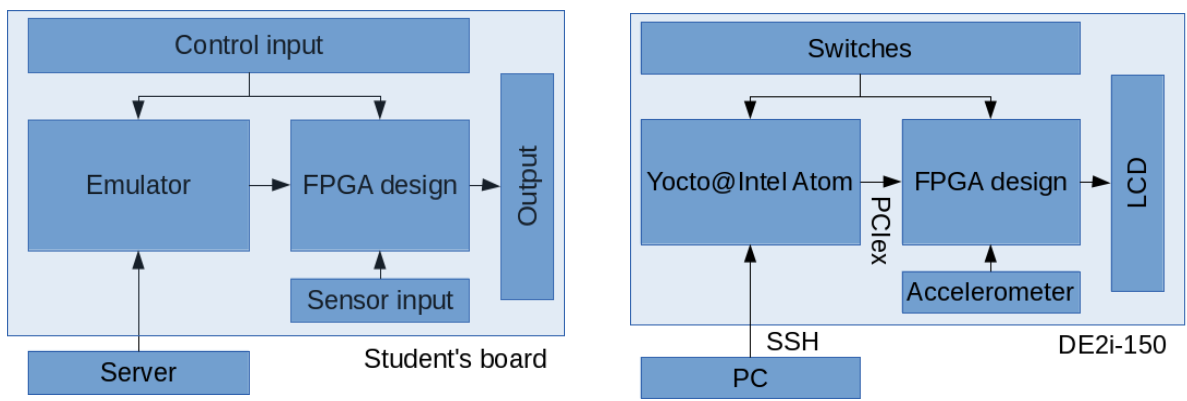

**Figure 2.** Generalized system design (a) and particular design on DE2i-150 (b)

Since attractor reconstruction is not usually implemented on FPGA devices, choosing to implement the AMIF-based method on the device in question is both a novelty and a proof of concept. The basis for auto-mutual information function implementation in Verilog that is going to be placed on the FPGA is the MATLAB code from [12]. It can be simplified generically to the following sequence of steps:

1. Converting the data into the 1-128 range
2. Making of histogram *A* for all data points except for the last $l_{max}$, where $l_{max}$ is the maximum lag, *m* points in total
3. Making histogram *B* for *m* data points starting at index (lag) *l*, where *l* changes between 1 and $l_{max}$ Making common two-dimensional histogram *AB* for sets *A* and *B*
4. For every value *v* in this common histogram, calculating $v \cdot \log_2 v/(v_A v_B)$ where $v_A$ and $v_B$ are corresponding values in histograms *A* and *B* and summing all these expressions. This is AMIF(*l*).

The procedure repeats until $l = l_{max}$. It is easily seen that this is indeed equivalent to the previously introduced formula (1). This procedure is shown graphically in Figure 3 as well.

This algorithm however needed to be converted in a real-time form, so that it does not wait for the whole data buffer to fill to perform the AMIF computation. This has led to the hardware design shown in Figure 4.

At every received sample from the sensor at the input, the algorithm performs updating the histograms and updating the value of the auto-mutual information function. This in practical terms means:

1. Update two numbers in histogram A: count in the bin the new sample belongs to is incremented and the count in the bin the sample that is now being removed from the calculation window is decremented.
2. Same goes for histograms B and AB, with a note that there will be $l$ of these.
3. The AMIF value is updated by updating the terms in it which have been influenced by the new sample (and the removal of the oldest one). This counts for two rows and two columns of AB histogram (the added sample column, the removed sample column, the delayed added sample row and the delayed removed sample row).

The exchange of information with the memory is limited to updating a small constant number of values and fetching a small number of rows and columns of histogram AB. The question of implementing the $l$ times repeated parts of the design is addressed later in the paper, together with time and space constraints.

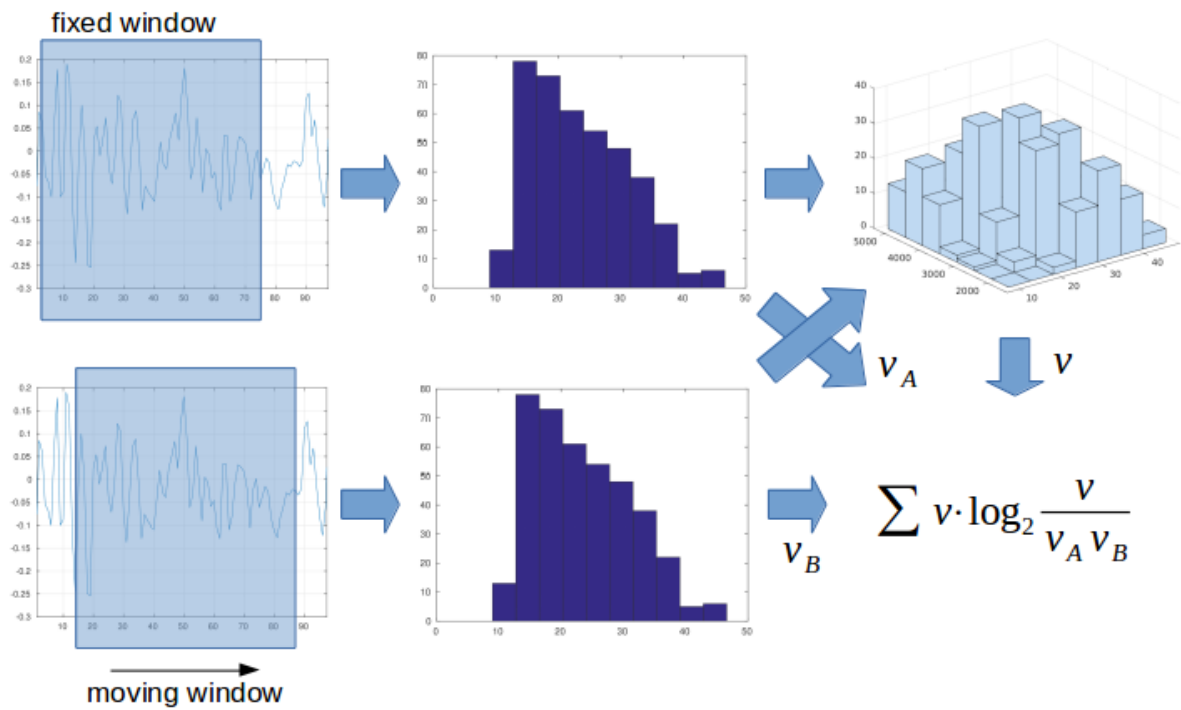

**Figure 3.** Graphical representation of AMIF calculation process

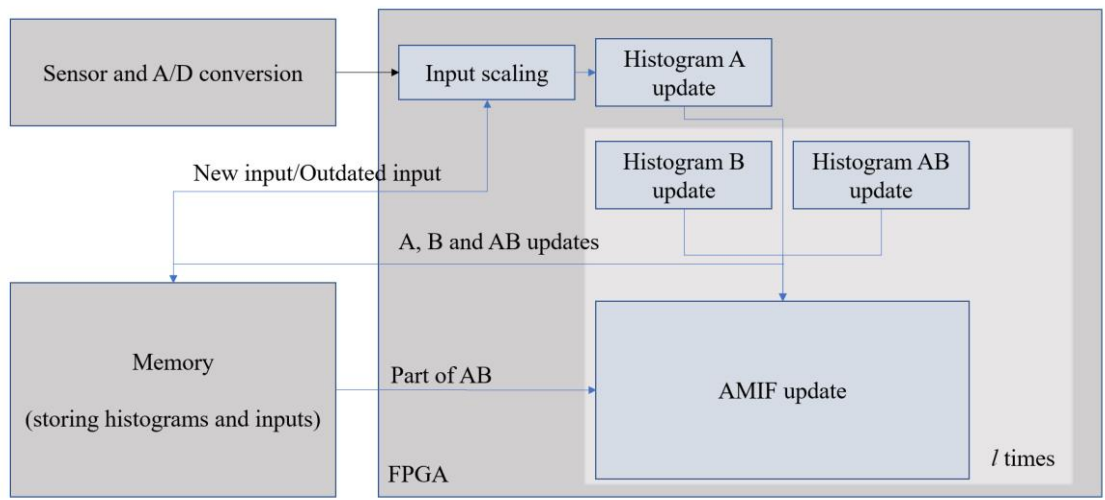

**Figure 4.** Hardware architecture of real-time AMIF calculation

Of course, other algorithms could have been implemented as well. The existence of Altera's megafunctions for FFT (Fast Fourier Transform) makes this procedure easier for the Fourier spectrum-based methods. Implementation of other methods may be a part of future work.

### 2.3. Artificial Aging

In the artificial motor aging data acquisition process, an induction motor was subject to chemical, thermal and electrical artificial aging compliant to the standard procedure.[†] The electrical discharge needed for fluting from the shaft to the bearing to cause bearing damage was induced in an experimental setup [17,18].

The fluting run of 30 minutes consisted of the 5 HP induction motor running with no load and with an external 27 A, 30V AC current to the shaft. In each cycle's end, the motor subject to aging was run on full load to measure its performance, recording data at a sampling frequency of 12 kHz. The motor was placed on a motor performance test platform. From the experimental setup seen in Figure 5, high-frequency data with a sampling frequency of 12 kHz was collected. In this case, we are working with eight 10-second time series (120,000 samples) representing motor condition from the healthy case (state 0) to the last working case (state 7).

---

[†] IEEE-117 (1974) IEEE Standard Test Procedure for Evaluation of Systems of Insulation Materials for Random-Wound AC Electric Machinery.

The data collected in this experiment was used in several studies on motor state detection based on vibration measurements [3, 18-22]. In this paper, we do not only aim to demonstrate the applicability of AMIF-based approach to age determination but also to give a wider perspective through implementing it in hardware and offering a framework for training and testing purposes. In terms of the relationship of the results presented here and results of other methods, a comparison between them can be found in [22]. The method proposed here has the advantage of perfectly separating the aged from the good condition: it does not give a monotonic indicator of deterioration, but a sharp threshold "good – aged". That makes it the best option in the binary classification option [22].

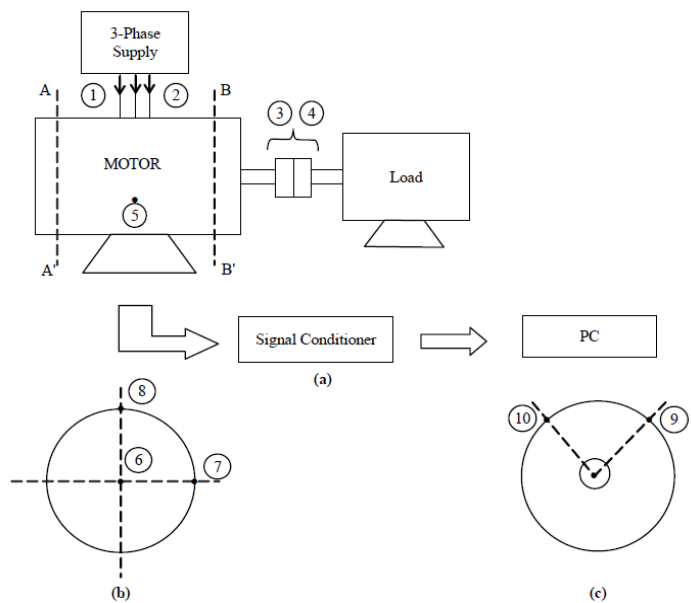

**Figure 5.** The experiment setup: (a) configuration, (b) and (c) cross-section AA' and BB' respectively. Numbers 1-4 denote electrical and mechanical measurement sensors, 6-10 accelerometers. [17,18]

## 3. Results

### 3.1. Attractor Reconstruction

Using the OpenTSTOOL, plots of auto-mutual information function for all the time series were obtained. Two characteristic plots (for time series 0 and time series 7) are shown in Figure 1. First minima for the first four signals are equal to 4, while the first minima for the second four signals are equal to one. These values are used for embedding dimension estimation. Plots of the dimension-determining Cao characteristics are omitted, but it is worth mentioning that the location of the first drop in the characteristic for any of the vibration signals is larger than 8. As [23] has shown that with 100,000 points only the dimension less than eight can be reconstructed, further reconstruction and Lyapunov exponent calculation are not feasible.

### 3.2. Automated Hardware Description

MATLAB's HDL Coder enables conversion of MATLAB m-files to VHDL/Verilog for most of Altera's FPGA chips, so the conversion of the auto-mutual information function code was conducted automatically. Before having the HDL Coder convert the m-file to Verilog, adjustments must be made on the available code. However, the code produced by the converter is additionally manually optimized and prepared for the particular hardware deployment, memory manipulation, etc.

To complete the standalone system, one only needs to provide the C-code to run on the Atom processor which would feed the data stream to the FPGA, together with a module to buffer the data. Since this Verilog code's verification has been conducted in MATLAB by using the testbench environment, there was no need for the development of the C-side, but it is conceptually clear.

The AMIF update part in Figure 4 is the only computationally demanding part of the design: it performs *4n* multiplications, *8n* logarithm calculations and *8n* additions where *n* is the number of sampling levels. For a single value of *l* this brought the design in our testing phase to 3 thousand logic elements used (the rest of the logic can be ignored for this analysis) for 128 sampling levels. Having 150,000 logic elements available in the development system we used, this meant the possibility of implementing all *l* repetitions of the AMIF updates (and corresponding histograms) in parallel: for 15 lag values it had us using less than 10% of the elements available.

Regarding the input frequency achievable, the design has no problem with frequencies up to 3 MHz. Since the sampling frequency of the data in our case was 12 kHz, we note that executing *l* repetitions could have been done on a single module in a loop without affecting real-time properties of the algorithm. For 15 lag values this approach lowers the effective frequency to 200 kHz, which is still an order of magnitude above the sampling frequency of the accelerometer data.

### 3.3. Motor age detection

Results of workbench tests on 512 samples from different parts of artificial motor aging signals are shown in Table 1. The testing was conducted by taking 15 windows of 512 samples from each of the artificial motor aging vibration series and leading it to the AMIF module with 32 sampling levels. These low figures both in terms of the number of samples and the number of sampling levels are taken to show how low can the optimization go, and still provide insightful results for condition monitoring. Smaller numbers here also mean simpler synthesis, as the histogram implementation has the complexity of *n* or *n²*, where *n* is the number of sampling levels. If we use the MATLAB

HDL Coder directly, for small values of $n$ this can directly be implemented on the FPGA, even without using the RAM chips.

**Table 1.** 15x8 testing on hardware implementation of AMIF

| Sample/series | 0 | 1 | 2 | 3 | 4 | 5 | 6 | 7 |
|---|---|---|---|---|---|---|---|---|
| Mean | 3.73 | 3.67 | 2.67 | 2.4 | 1 | 1.87 | 1.73 | 1 |
| Std. dev. | 0.7 | 1 | 0.7 | 1.2 | 0 | 1.2 | 1.4 | 0 |

**Table 2.** 10 x 8 testing of AMIF on non-artificial aging

| Day | 1 | 2 | 3 | 4 | 5 | 6 | 7 | 8 |
|---|---|---|---|---|---|---|---|---|
| Mean | 2.7 | 2.9 | 2.8 | 2.5 | 2.7 | 2.5 | 1.9 | 1.9 |
| Std. dev. | 0.7 | 0.7 | 1.3 | 0.5 | 1 | 1.1 | 0.7 | 1 |

Tests ran on this system also include: the 120,000 samples (whole time series) test, 16,000 samples test, both run on the vibration time series; surrogate data test on 120,000 samples phase shuffled vibration data; validation of the algorithm on 8192 samples of Rossler chaotic time series.

Another test was conducted to confirm the applicability of this system, following the example set in [19], where data from non-artificial aging process was used to confirm the efficiency of an algorithm designed on artificial motor aging data. The experimental data were collected in a study[‡] [24] where a motor was running for 8 days, collecting 1-second long vibration data at 10-minute intervals before breaking down due to bearing failure. In this study, just like in [19] data from the accelerometer placed on the later failed bearing was used (Figure 6). Ten 1 second samples were randomly picked for each day and algorithm was applied to them with results shown in Table 2.

---

[‡] Lee J, Qiu H, Yu G, Lin J. Rexnord Technical Services. Bearing data set. IMS, University of Cincinnati, NASA Ames Prognostics Data Repository, http://ti.arc.nasa.gov/project/prognostic-data-repository, 2007.

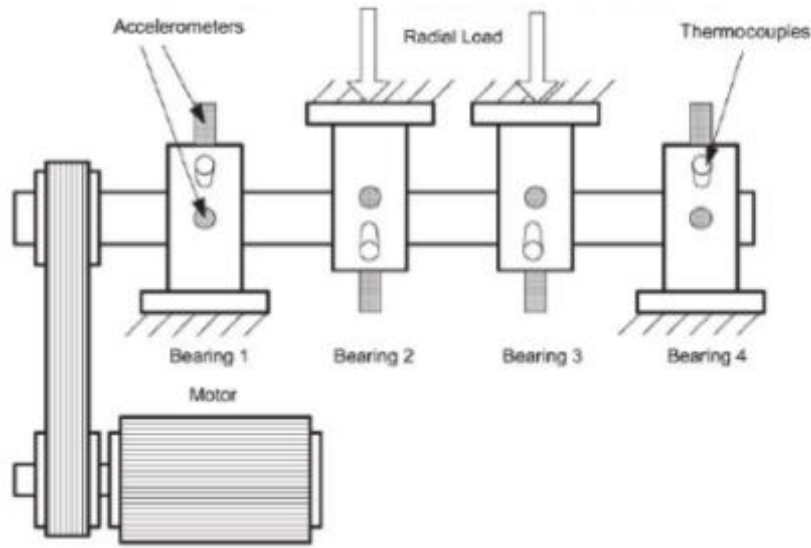

**Figure 6.** Setup in non-artificial motor aging case [24]

Results obtained in this section indicate a relationship between the motor age and the index of the first auto-mutual information function minimum. This link will be discussed in detail in the following section.

## 4. Discussion

The important result of this research is the delay time as extracted from auto-mutual information function (first minimum). The first four time series (representing the first half of aging process) have $\tau = 4$, while the second four (second half of aging process) have $\tau = 1$. This is a discriminatory feature for determining roughly the motor age, therefore applicable for maintenance. An interesting fact worth investigating further is that the same is found for surrogate data as well (applying the surrogate algorithm described in [25]), meaning that the "color" of the noise-vibration is dominant in the sense of this feature. Still, as it has been seen before [20], Hurst exponent as another signal feature based on noise color does not follow a similar pattern.

How is the auto-mutual information function related to motor health? With the change of motor state, amplitude and frequency of motor vibrations change. An energy shift towards higher frequencies due to inner fluting faults has been observed before [17]. With this change in frequency and amplitude, it takes fewer samples (less time) for the auto-mutual information function to reach the minimum, and hence the index value decreases. It has been shown before, however, that observing only frequency or only amplitude is not as efficient [17].

The second result is the conclusion that no low-dimensional chaos appears in any of these vibration signals, leaving two options open: high-dimensional chaos and stochastic nature, and the high-dimensional chaos option is unverifiable based on the theoretical results in [19].

Results obtained from the hardware implementation of AMIF somewhat differ from the results from MATLAB for 120,000 samples. Namely, results show a slightly anomalous behavior for time series 4 (which has been reported before, cf. [21]). This is not crucial, as the 512-point implementation was only for testing purposes. The statistics also show the irrelevance of standard deviation for these calculations, as seen in the table. If the number of samples is increased to 16,000, no deviation exists at all, and all 16,000 samples long time series for the particular motor state return the same result.

It is worth mentioning that the system is validated with the Rossler data with the known first minimum of AMIF at sample no. 14 according to [8]. The algorithm returned 14 as a result even at low sample counts, therefore verifying itself. This implementation allows age differentiation at even a small number of data points (512, which is collected in less than 0.05 seconds). That suggests that the working hypothesis that a simple FPGA system may be designed for condition monitoring of motors subject to artificial motor aging is in fact correct, where a function previously not implemented in hardware got its hardware version. While it is true that this algorithm may be implemented efficiently in digital signal processors and other processors, a hardware implementation may improve performance and ensure it is not the processing bottleneck. Quick acquisition and processing in this case (under 0.1s) enable real-time work.

The test on other motor data (non-artificial aging) with results shown in Table 2 indicates an abrupt change for day 7 and 8, the same behavior seen in [19] and in the original research [24] caused by the bearing fault. This pattern matches what has been seen in case of artificial motor bearing (i.e. Table 1) and hence supports the possibility of application of this method in motor condition monitoring.

The artificial motor aging dataset has been extensively used ever since the data was collected [17], with some results presented in cited publications [3, 18-22]. Similar to the results of Lyapunov exponent application in [3] or Hurst exponent application in [20], this method shows a clear difference between the results on data set 0 and data set 7. However, it also gives a sharp bound in between, at which the state of the motor significantly changes. On the other hand, there are methods that can measure motor age in a monotonic manner [19, 21] which was not the goal of this method. As mentioned earlier a comparison of data provided by all these methods in terms of feature ranking can be found in [22].

## 5. Conclusions

We have here presented a hardware implementation of auto-mutual information function serving as a condition monitoring tool for electrical motors and indicating the motor state in both artificial and non-artificial motor aging processes. From a wider perspective, a flexible framework for condition monitoring and education purposes based on a customizable hardware architecture was introduced with the auto-mutual information function as a proof of concept.

While the attractor reconstruction in the case of motor vibration analysis did not prove to be particularly meaningful, the work done here in hardwarization of auto-mutual information function calculation algorithm could serve as a basis for a hardware-based attractor reconstruction module which would be an effective tool in real-time, practical dynamical systems analysis.

Future work in this area may also involve testing the platform on diverse types of motors, implementing different algorithms in the hardware, as well as using the platform in the classroom for various courses. Furthermore, development of custom hardware, a customized version of the general-purpose FPGA board as seen in this letter is a possible next step in this research. Application of auto-mutual information function in the industry is yet to be seen.

**Acknowledgement**

This work was supported in part by the Altera Inc. University Program grants BR8133 and LR1300 (2014).